\documentclass[12pt,a4paper,notitlepage,headings=normal, DIV9, parskip=half-, headinclude]{scrartcl}

\usepackage{bm}
\usepackage{graphicx}
\usepackage{amsmath}
\usepackage{amssymb}
\usepackage[superscript,nocompress]{cite}
\usepackage[left=2cm,right=2cm,top=2cm,bottom=3cm]{geometry}
\usepackage[numbers]{natbib}

\usepackage[utf8]{inputenc}

\usepackage[plainpages=false,pdfpagelabels,colorlinks=true,linkcolor=blue,urlcolor=blue,citecolor=blue,pdfduplex=DuplexFlipLongEdge]{hyperref}	
\begin{document}
\title{Comment on ``Consistent thermostatistics forbids negative
absolute temperatures'' }
\date{}
\maketitle

{\vspace{-2.5cm}\centering \large Ulrich Schneider$^{*,1}$, Stephan Mandt$^2$, Akos Rapp$^3$, Simon Braun$^1$, Hendrik Weimer$^3$, Immanuel Bloch$^1$, Achim Rosch$^4$\\
\vspace{5mm}\small $^*$ ulrich.schneider@lmu.de
\small $^1$ LMU \& MPQ München, $^2$ Princeton Center for Complex Materials, Princeton University, $^3$  Institute for Theoretical Physics, Universität Hannover, $^4$ Institute for Theoretical Physics, Universität zu Köln \par }

In their paper~\cite{Dunkel2013}, Dunkel and Hilbert argue that negative absolute temperatures, a well-established thermodynamic concept, are inconsistent with thermodynamics. They claim to ``{\em prove that all previous negative temperature claims and their implications are invalid as they arise from the use of an entropy definition that is inconsistent both mathematically and thermodynamically}''. Here we point out that, on the contrary, negative temperatures are not only thermodynamically consistent for systems with bounded spectra ($|\langle\hat{H}\rangle|/N\leq\epsilon$ with $N$ the number of constituents and $\epsilon <\infty$), such as spin systems or Hubbard models, but are in fact necessary to describe equilibrated states of these systems in the high-energy regime, where the density of states, and thereby the conventional entropy, drops as a function of energy. Recently, we have successfully predicted~\cite{Rapp2010} and experimentally observed~\cite{Braun2013} negative absolute temperatures using ultracold atoms in optical lattices. 

To show how thermodynamically consistent negative temperatures can be used for \emph{bounded} Hamiltonians, we start from conventional (not inverted) thermal states of a given bounded Hamiltonian $\hat{H}$, described by density matrices $\rho$, entropies $S$ and positive absolute temperatures $T\ge 0$. For the inverted Hamiltonian $\hat{H}'=-\hat{H}$ with the same density matrix $\rho'= \rho$ we define the entropy $S'$ according to $S'=S$. As $E'=-E$, it directly follows that the resulting temperature has to be negative, since $T'=(d S'/d E')^{-1}=- (d S/d E)^{-1}=-T$. With this definition, namely $S'=S$, all thermodynamic relations are trivially fulfilled for the $\hat{H}'$ system if they have been fulfilled for $\hat{H}$. The simplest example for such a construction is a canonical ensemble, where
 \[\rho=e^{-\hat{H}/(k_B T)}/Z=e^{-(-\hat{H})/(- k_B T)}/Z=e^{-\hat{H}'/(k_B T')}/Z=\rho'.\] 
Combined with the usual definition of entropy, $S=-k_B {\rm Tr}( \rho \ln \rho)$, the canonical ensemble therefore fulfills all conditions demanded by Dunkel and Hilbert for a consistent thermodynamics (see Eq. (4-6) in~\cite{Dunkel2013}), as can be checked straightforwardly and is acknowledged by Dunkel and Hilbert in \cite{Dunkel2014}. 
Thus, the only remaining issue is the question of the equivalence of ensembles for inverted systems. 
In the above situation with $\hat{H}'=-\hat{H}$ and $\rho'= \rho$, however, any proof given for the conventional case of $T>0$ and Hamiltonian $\hat{H}$ can be directly translated to $T'<0$ for Hamiltonian $\hat{H}'$. 
Therefore the main claim of Ref.~\cite{Dunkel2013} is not valid.

In the remainder of this comment, we discuss advantages and disadvantages of the alternative point of view advocated by Dunkel and Hilbert in~\cite{Dunkel2013}: 
The authors choose a non-standard definition of the entropy of microcanonical ensembles using the so-called Hertz entropy (unfortunately named Gibbs entropy in Ref.~\cite{Dunkel2013}), $S_G=k_B \ln \Omega$, where $\Omega= {\rm Tr}\ \Theta(E-\hat{H})$ counts all states up to energy $E$, starting from the bottom of the spectrum. With this definition, the entropy by construction always increases monotonically as a function of energy and therefore temperatures calculated from this definition are always positive, $T_G=(d S_G/d E)^{-1}>0$. 
As pointed out by Dunkel and Hilbert, $S_G$ and $T_G$  have some advantages in case that one does insist on defining temperatures for {\em small microcanonical} systems. 
While temperature is a very powerful and useful concept in the thermodynamic limit
and for small (canonical) ensembles in contact with a thermal bath, its usefulness for \textit{small microcanonical} systems is limited:
Dunkel and Hilbert correctly point out that for such systems the second part of their Eq.\ (6) in~\cite{Dunkel2013} is not valid when the standard Boltzmann entropy is used.  This does, however, by no means imply that standard microcanonical thermodynamics is ``thermodynamically inconsistent'', it is rather one of several problems that can arise when one tries to define thermodynamics for \textit{small} microcanonical systems. For example, an infinitesimal coupling of two small microcanonical systems with formally identical initial $T$ typically induces a heat flow changing $T$~\cite{Gibbs,Frenkel2014}. 

It is furthermore interesting to note that, for systems with bounded spectra, one can equally well consider an inverted version of the microcanonical Hertz entropy by defining $S_{G'}=k_B \ln \Omega'$ with $\Omega'= {\rm Tr}\ \Theta(-E+\hat{H})$, in which the number of states is counted from the top instead of the bottom of the spectrum. The resulting temperature $T_{G'}$ is always negative and has, in contrast to $T_G$, the correct thermodynamic limit for inverted states.

To show the disadvantages of the postulates used by Dunkel and Hilbert, in the following we list a number of severe problems of their approach: 

(i) The entropy $S_G$ is unphysical in the sense that it cannot be computed from  the density matrix alone. It furthermore depends on states that are energetically inaccessible to the system. 

(ii) $S_G$ cannot be connected to foundational concepts of modern statistical physics based on information theory: The highest Hertz entropy would, for example, be associated to the ensemble at the highest possible energy, which typically consists of only one non-degenerate state such as, e.g., all spins up in a spin system. The highest $S_G$ therefore corresponds to a pure state with zero entropy in conventional approaches. 

(iii) In inverted situations the thermodynamic limit  of the approach of Dunkel and Hilbert is completely ill defined~\cite{Vilar2014}.  

(iv) $S_G$ violates the second law of thermodynamics in the formulation that entropy cannot decrease in an isolated system. Consider a many-body system with a unique highest excited state: the microcanonical ensemble at this energy has zero conventional entropy $S$ but maximal $S_G$. If one perturbs this system by, e.g., a quantum quench, it can undergo transitions to lower energy states such that the density matrix becomes mixed and $S$ increases. In the formalism of Dunkel and Hilbert, however, this would correspond to a decreasing entropy $S_G$~\cite{Campisi2008}. 

(v) $T_G$ violates that 
heat should always flow from the hotter (smaller $1/T$) into the colder (larger $1/T$) system. Consider a finite microcanonical system with bounded spectrum and inverted population, where the energy is larger than that of the $T=\infty$ canonical ensemble. The use of $S_G$ ascribes a finite positive $T_G$ to this state. Next we bring this system into weak thermal contact with an unbounded and infinitely large thermal bath at $T=T_G=\infty$. Once the system has equilibrated with the bath, it is described by a canonical density matrix at $T=\infty$ and has thereby lowered its energy. Therefore heat has flowed from the inverted system with $T_G< \infty$ into the $T=\infty$ bath.

(vi)  $T_G$ is inconsistent with  $T$ defined for canonical ensembles. Consider the microcanonical ensemble
of an infinitely large system with inverted population. 
The reduced density matrix of a finite subsystem can accurately be described by a canonical ensemble with $T<0$ while $T_G >0$. In contrast, microcanonical and canonical temperatures coincide when the usual 
Boltzmann entropy is used.

Although the vast majority of physical systems have no upper bounds in energy (e.g.\ the kinetic energy $p^2/2m$ is unbounded) and can therefore sustain only positive $T$, for systems with bounded spectra  negative absolute temperatures are a well-established concept, which is not only consistent with thermodynamics, but unavoidable for a consistent description of the thermal equilibrium of inverted populations. 

\bibliographystyle{plain}

\end{document}